# Understanding Phonon Properties in Isoreticular Metal-Organic Frameworks from First Principles

*Tomas Kamencek[1,2], Natalia Bedoya-Martínez[3], and Egbert Zojer[1]*

[1]*Institute of Solid State Physics, Graz University of Technology, NAWI Graz, Petersgasse 16, 8010 Graz, Austria*

[2]*Institute of Physical and Theoretical Chemsitry, Graz University of Technology, NAWI Graz, Stremayrgasse 9, 8010 Graz, Austria*

[3]*Materials Center Leoben, Roseggerstraße 12, 8700 Leoben, Austria*

**ABSTRACT:** *Metal-organic frameworks (MOFs) are crystalline materials consisting of metal centers and organic linkers forming open and porous structures. They have been extensively studied because of various possible applications exploiting their large amount of internal surface area. Phonon properties of MOFs are, however, still largely unexplored, despite their relevance for thermal and electrical conductivities, thermal expansion, and mechanical properties. Here, we use quantum-mechanical simulations to provide an in-depth analysis of the phonon properties of isoreticular MOFs. We consider phonon band structures, spatial confinements of modes, projected densities of states, and group velocity distributions. Additionally, the character of selected modes is discussed based on real-space displacements, and we address how phonon properties of MOFs change when their constituents are altered, in terms of mass and spatial extent, bonding structure, etc. We find that more complex linkers shift the spectral weight of the phonon density of states toward higher frequencies, while increasing the mass of the metal atoms in the nodes has the opposite effect. As a consequence of the high porosity of MOFs, we observe a particularly pronounced polarization dependence of the dispersion of acoustic phonons with rather high group velocities for longitudinal acoustic modes (around 6000 m s−1 in the long-wavelength limit). Interestingly, also for several optical phonon modes group velocities amounting to several thousand meters per second are obtained. For heterogeneous systems such as MOFs, correlating group velocities and the displacement of modes is particularly relevant. Here we find that high group velocities are generally associated with delocalized vibrations, while the inverse correlation does not necessarily hold. To quantify anharmonicities, we calculate mode Grüneisen parameters, which we find to be significant only for phonons with frequencies below ∼3 THz. The presented results provide the foundations for an in-depth understanding of the vibrational properties of MOFs and, therefore, pave the way for a future rational design of systems with well-defined phonon properties.*

## I. INTRODUCTION

Metal-organic frameworks (MOFs) are highly porous structures consisting of metal(-oxide) clusters (the nodes), which are connected by organic molecules (the linkers). The huge variety of organic chemistry renders this material class highly interesting for a number of applications exploiting the high porosity of these structures (e.g. catalysis [1],[2] and storage and separation of gases [3]–[6]).

To unleash the full potential of MOFs for applications in functional devices, the correlation between their structures and their physical and chemical properties must be understood. Many of these properties strongly depend on vibrations and, consequently, for periodic systems such as MOFs on their phonon band structures: For example, phonon contribution to the free energy are crucial for reliably predicting phase stability [7],[8]; in the long-wavelength limit, the slopes of the acoustic branches determine the elastic behavior of the crystals impacting mechanical stability as a crucial factor for the (oriented) integration of MOFs into devices [9]; the coupling between phonons and electrons determines charge transport; the (negative) thermal expansion observed in certain MOFs can be associated with specific phonon bands[10]–[15]; also thermal energy in (pristine) MOFs is transported by phonons, which makes heat dissipation or the thermoelectric figure of merit strongly dependent on (anharmonic) phonon properties.

Finally, as far as $\Gamma$-point phonons are concerned, they determine the vibrational spectra of MOFs, which can be used to determine their composition, and to highlight differences in their structural properties [16]–[20].

All these aspects call for a detailed understanding of phonon properties of MOFs, which is, however, still in its infancy [14],[21]. The situation is further complicated by the enormous number of possible MOF structures, with the potential that some of them have outstanding properties.

In this work we therefore employ quantum-mechanical simulations to develop systematic relationships between the phonon properties of MOFs and the nature of their building blocks. As we are at a very early stage of this quest, we focus on the (harmonic) vibrations of cubic isoreticular MOFs (IRMOFs), i.e., structures with identical linkers along the edges of the cubic repeat units, see FIG. 1(a). In particular, we analyze the relation between the MOF structure (varying linkers and nodes) and quantities such as the (projected) densities of states, the heat capacity, phonon bands, the degree of localization of specific vibrations, group velocities, and mode Grüneisen parameters. In this way we develop an in-depth understanding of phonon properties of MOFs, laying the foundations for future phonon-engineered systems with properties optimized for specific applications.

The paper is organized as follows: We start with an introduction of the studied systems, followed by a detailed description and benchmarking of the applied methodology. In the subsequent section, a general discussion of the phonon density of states and the resulting heat capacity will be presented, followed by a detailed analysis of the phonon bands, group-velocity distributions, and mode localization. Additionally, we briefly discuss anharmonic properties characterized by (mode) Grüneisen parameters. We mostly focus on the low-frequency regime [below 3 THz (100 cm$^{-1}$)], which is particularly relevant for the materials' thermal properties.

## II. STUDIED SYSTEMS

The structure of the considered IRMOFs consists of (comparably heavy) nodes formed by four XO$_4$ tetrahedra (with X representing a metal), which share one oxygen atom as a common corner. Neighboring nodes are connected by organic molecules via bonds to two oxygen atoms on each side, which results in a cubic pore [see FIG. 1(a)]. Notably, two neighboring nodes are each other's mirror images. Thus, two nodes and their respective linkers form the crystallographic basis, which is replicated on a face-centered cubic (fcc) lattice.

The connections in FIG. 1(a) indicated by blue lines represent one of the several possible linker molecules. The chemical structures of the three linkers considered in this study are shown in FIG. 1(b-d): IRMOF-1 [with the linker being shown in FIG. 1(b)] is the most studied IRMOF and, thus, can be viewed as the prototypical IRMOF system. For the Zn-based variant even its phonon band structure has been studied in the context of (negative) thermal expansion [14]. The linker shown in FIG. 1(c) is the shortest possible dicarboxylic acid linker. This renders the system fundamentally relevant. As the resulting MOF lacks a common name, it will from now on be referred to as IRMOF-130. The small number of atoms in the linker significantly reduces the number of optical phonon modes, which will simplify the following discussion. IRMOF-130 has not been experimentally realized yet, but its Zn- analog has already been the subject of theoretical studies [22]. As a third system, we will study IRMOF-14, which is formed by the rather heavy and more extended aromatic dicarboxylic acid linker shown in FIG. 1(d). Being the largest system studied here, it contains 190 atoms in the primitive unit cell, which results in a particularly rich phonon spectrum comprising in total 570 different phonon bands. Notably, IRMOF-14 is characterized by a rather large moment of inertia for rotation around the linker, which ought to profoundly reduce the frequencies associated with torsional vibrations.

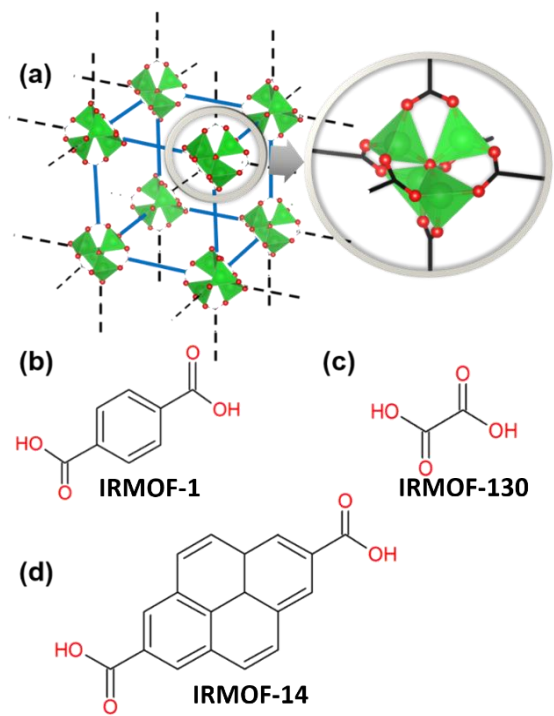

FIG. 1. (a) Exemplary IRMOF geometry in which the linkers have been replaced by blue lines. The nodes, which are shown in the circular inset, consist of four $XO_4$ (X = Mg, Ca, …) tetrahedra, which share one oxygen atom. Coloring scheme: X green, O red. (b)-(d): Chemical structures of the studied linker molecules (before deprotonation), which are located between the nodes of the IRMOFs.

Concerning the metal atoms in the node, the Zn-based analogs of IRMOF-1 and IRMOF-14 are the most commonly investigated structures [14],[23]–[25]. Here, we focus on the less common Mg- (as reference system) and Ca-based systems, which have also been considered in several studies [25]–[28]. The reason for choosing Mg rather than Zn is mostly technical, as detailed in the Supplemental Material [29]. For the sake of gaining fundamental insights, the choice of metal atom is, however, rather inconsequential, as the bonding chemistry between metal nodes and linkers has been found to be similar for metal atoms with the same formal charge (oxidation number) [26],[27]. Thus, varying the metal (here, exchanging Mg with Ca) is expected to primarily impact the phonons in a MOF through the different masses of the nodes.

## III. METHODOLOGY

### A. Density functional theory and density functional tight binding calculations

Density functional theory (DFT) [30],[31] calculations were carried out with the *VASP* code [32]–[35] (version 5.4.1), using the PBE functional [36] and employing the projector-augmented wave method [37], together with the recommended standard potentials [38] specified in the Supplemental Material [29]. There we provide also additional details regarding specific numerical settings and the converged k-meshes and plane wave energy cutoffs for all systems. The geometries and unit cell parameters of all MOFs were optimized until the maximum force component was below 0.5 meV/Å (see Supplemental Material [29]). In these optimizations, the space group was fixed as $Fm\bar{3}m$.

The above-described DFT calculations were performed for the primitive unit cells of all studied systems. For the extended supercells necessary for studying phonon dispersion relations (see below), such calculations, however, turned out to be prohibitively expensive. Therefore, when calculating off-Γ phonons, the computationally more affordable density functional tight binding (DFTB) approach was used [39]–[42]. This method, despite its approximate character, provides a good compromise between accuracy and efficiency. It has been shown that it yields structural [43] and thermal transport properties [21] of MOFs comparable to the DFT level of theory. The computations based on the self-consistent charge (SCC) DFTB approach were performed using the *DFTB+* [44] package (version 18.1), and the *3ob-3-1* Slater-Koster files including the *3ob:freq-1-2* extension for obtaining more accurate vibrational properties [45]–[47].

A detailed description of the DFTB simulation parameters can be found in the Supplemental Material [29]. To ensure high-quality geometries, the maximum residual forces in the geometry optimizations were set to $10^{-5}$ meV/Å.

All results reported here were obtained without *a-posteriori* van der Waals (vdW) corrections. This is justified, as the systems of interest consist of covalently and coordinatively bonded constituents. Indeed, tests on IRMOF-130(Mg) show that including van der Waals interactions hardly changes the phonon frequencies (see Supplemental Material [29]; the observed RMS deviation between calculations with and without van der Waals corrections was below 0.05 THz, i.e., below 1.5 cm$^{-1}$).

### B. Phonon properties

Phonons were calculated in reciprocal space by means of lattice dynamics (LD) within the harmonic approximation to the potential energy surface and employing the finite displacement method as implemented in the PHONOPY [48] code. A displacement amplitude of 0.01 Å was used, which is motivated by tests on IRMOF-130(Mg) (see Supplemental Material [29]). The $\Gamma$–phonons for the comparison between DFT and DFTB were obtained employing primitive unit cells, while converged phonon band structures were calculated for 2x2x2 supercells of the conventional fcc cells for all considered materials (see Supplemental Material [29]). These supercells contain 32 primitive cells.

In order to discuss anharmonicities, (mode) Grüneisen parameters were calculated with the PHONOPY code from phonon band structures at slightly enhanced and reduced volumes (~±0.3%).

### C. Analyzed quantities

Phonon densities of states, band structures, group velocities, and mode participation ratios were used to analyze the relationship between the structural and vibrational properties of MOFs.

Vibrational modes were characterized and compared among systems on the basis of the associated displacement patterns. Technically, this was mostly done by a visual inspection of the corresponding animations considering the symmetries of the vibrations. Additionally, plots of the displacement patters were created with VESTA 3[49].

The reported densities of states (DOSs) were normalized such that integrating over them yields 3N, where N is the number of atoms in the primitive unit cell.

Less common quantities used in the later discussion are briefly described in the following.

#### 1. Frequency and group velocity resolved density of states

For phonon transport, the group velocities of the phonons are amongst the most relevant parameters, as they determine the time scales on which phonons propagate and exchange energy. Unfortunately, the sheer number of bands in MOFs makes an in-depth discussion of each of them together with the associated group velocities virtually impossible. A strategy for still being able to discuss trends in band dispersions and group velocities relies on calculating the density of states resolved not only with respect to the angular frequency ω, but also with respect to the norm of the group velocity vector, $\|v_g\|$. We define this 2D density function in analogy to the common definition of the DOS as

$$p(\omega, \|v_g\|) = \frac{1}{3N\,N_q} \sum_{q,n}^{N_q,3N} \delta(\omega - \omega_{q,n}) \delta\left(\|v_g\| - \|v_g\|_{q,n}\right) \quad (1)$$

The prefactor in Eq. (1) contains the number of bands (3N) and the number of wavevectors $N_q$ used in the discrete sampling of the phonon bands in the first Brillouin zone (1BZ). In this way, the density function p is normalized to 1. The 2D density function shows, how frequently certain values of the group velocities occur at a given frequency. In practice, the δ distributions are replaced by Gaussian or Lorentzian functions with a finite width. Here, widths of 0.2 THz and 0.2 THzÅ were used for the (projected) DOS and for the group velocity resolved DOS. In both cases, for sampling the 1BZ a uniform *q* mesh containing 20x20x20 points was applied.

### 2. Mode participation ratios

In simple systems and textbook phonon models, bands with low dispersion are often associated with localized modes. The reason for that is that if only some of the atoms in the unit cell move, the face shift of the displacements in successive unit cells (expressed by **q**) has only a minor impact on the mode frequency. In MOFs, such localization effects ought to be particularly pronounced due to their heterogeneous nature caused by the presence of heavy as well as light atoms. Consequently, one would expect small group velocities for localized modes. To test this hypothesis, we will correlate the group velocities and mode localizations for the MOFs studied in this work.

A quantitative assessment of the degree of localization of a phonon can be achieved via the so-called participation ratio (PR) [50]–[52]. It is defined as

$$PR_{q,n} = \frac{\left(\sum_{\alpha=1}^{N} \frac{|e_{q,n}^\alpha|^2}{m_\alpha}\right)^2}{N \cdot \left(\sum_{\alpha=1}^{N} \frac{|e_{q,n}^\alpha|^4}{m_\alpha^2}\right)} \quad (2).$$

Here, **q** and n denote the wave vector and band index, N is the number of atoms in the unit cell, $e^\alpha_{q,n}$ are the three components of the vibrational eigenvectors of atom α, and $m_\alpha$ is its mass. It is easy to show from Eq. (2) that when the PR is on the order of 1/N, the mode is considered to be highly localized (only a few atoms move), whereas participation ratios on the order of 1 mean that the mode is highly delocalized (essentially all atoms in the unit cell move with the same amplitude). Since the eigenvectors depend on **q**, the participation ratio can vary significantly within a band.

## IV. RESULTS AND DISCUSSION

### A. Validation of the DFTB approach

As a first step, it is important to benchmark the accuracy of the DFTB calculations against the more reliable DFT results. This is done for the unit cell parameters and for the vibrations at the Γ-point.

**TABLE I: Comparison of the optimized lattice constants, *a*, obtained within DFT (VASP) and DFTB (DFTB+). Additionally, the root-mean-square deviations between the frequencies for modes up to 5 THz ($RMS_{\leq 5\,THz}$) and for all modes ($RMS_{full}$) are shown.**

| IRMOF | 1(Mg) | 130(Mg) | 130(Ca) | 14(Mg) |
|---|---|---|---|---|
| $a_{DFT}$ / Å | 13.10 | 8.82 | 9.62 | 17.35 |
| $a_{DFTB}$ / Å | 13.16 | 8.84 | 9.46 | 17.45 |
| $\Delta a/a_{DFT}$ / % | 0.46 | 0.23 | -1.66 | 0.59 |
| $RMS_{\leq 5\,THz}$ / THz (cm$^{-1}$) | 0.12 (4) | 0.26 (8) | 0.43 (14) | 0.13 (4) |
| $RMS_{full}$ / THz (cm$^{-1}$) | 1.29 (43) | 1.36 (45) | 1.40 (46) | 1.19 (39) |

Table I shows that for the Mg-based MOFs the equilibrium lattice constants obtained with the two approaches deviate by less than 0.6 %.

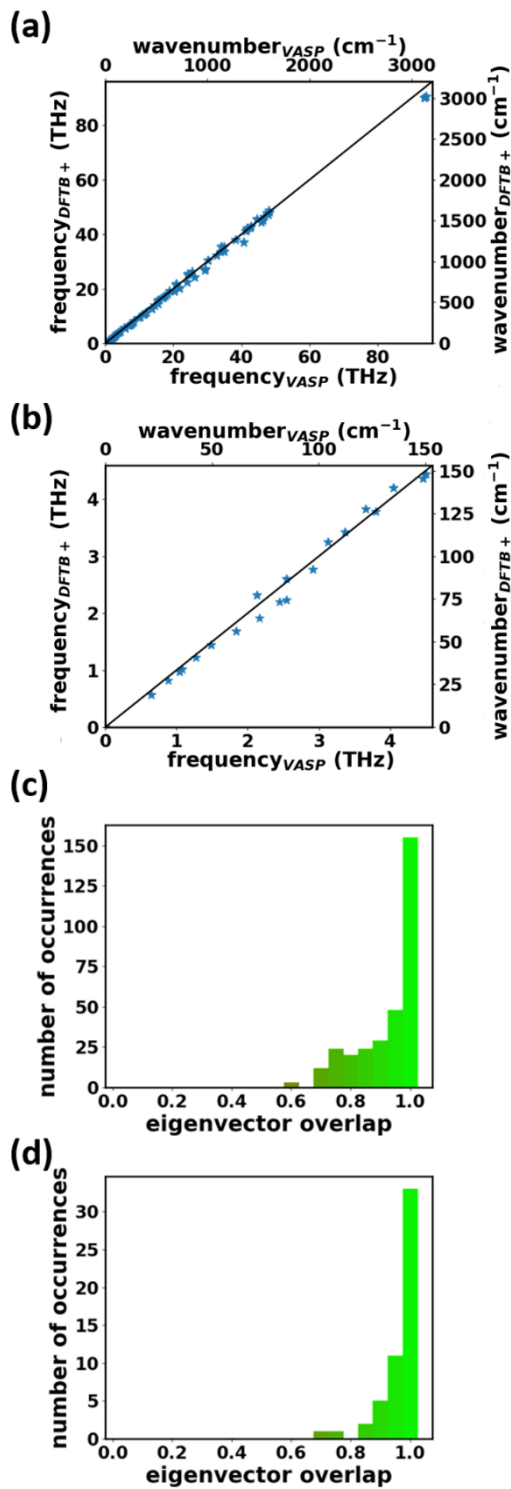

**FIG. 2.** Scatter plots of the Γ frequencies obtained with VASP and DFTB+ for IRMOF-1(Mg) in (a) the full frequency range and (b) for a reduced range comprising modes up to 5 THz. Panels (c) and (d) show histograms of the scalar products between eigenvectors obtained with VASP and DFTB+ for (c) all modes and (d) modes with frequencies $\leq$ 5 THz.

Also for the Ca-based MOF, the observed deviations (< 2 %) are at an acceptable level.

Moreover, the agreement between the frequencies at the Γ-point calculated with DFT and DFTB is satisfactory, as can be seen in FIG. 2(a) and (b) for IRMOF-1(Mg). Equivalent plots for IRMOF-130(Mg/Ca) and IRMOF-14(Mg) can be found in the Supplemental Material [29].

The deviations are clearly larger in the high-frequency than in the low-frequency region, which is confirmed by the corresponding RMS values in Table I. The distinction between the two spectral ranges is relevant insofar as for physical properties derived from the phonons usually the lower-lying modes are more important, as they are the ones that are thermally occupied.

As an additional benchmark, the eigenvectors of the dynamical matrix are compared between DFT and DFTB. As the dynamical matrix is Hermitian, its eigenvectors form an orthonormal basis, which allows an analysis of pairwise scalar products between them. The majority of the scalar products amounts to ~1 (perfect agreement) or to slightly smaller values [see FIG. 2(c) and (d)]. Again, a better overall agreement is obtained at smaller frequencies.

To investigate the situation away from the Γ point, we also calculated the frequencies at L employing VASP and a suitable supercell extended only along one lattice vector for IRMOF-130(Mg). A direct comparison with the DFTB data can be found in the Supplemental Material [29]. This comparison provides some interesting insights: for the acoustic phonons, both methodologies yield virtually identical results; this is also observed for some of the optical bands, while others appear mostly rigidly shifted between VASP and DFTB+. In some cases, however, also the band widths between the Γ and the L points change significantly. This suggests that at this stage calculated phonon band structures of MOFs provide mostly qualitative insights. For a fully quantitative validation of the results it would be extremely important to have access to suitable experimental data.

## B. Impact of structural variations on the phonon density of states

The densities of states projected onto the various atomic species are the ideal starting point for the discussion of the phonon properties of the MOFs introduced in Sec. II. They allow an assessment of the frequency ranges in which phonon modes exist in each of the materials. Moreover, they show, at which frequencies certain types of atoms primarily contribute to the oscillations. The projected DOSs (PDOSs) of IRMOF-130(Mg/Ca), IRMOF-1(Mg), and IRMOF-14(Mg) are shown in FIG. 3(a) for the entire frequency range. A zoom into the frequency region up to 20 THz is shown in FIG. 3(b). The contributions from different atom types are plotted in a stacked manner, such that the outermost contour corresponds to the total DOS. The oxygens and the metal atoms are defined to form the node of the MOF. Their contributions to the PDOS are hatched to emphasize the different roles of linkers and nodes. Note that this separation in node and linker is not unique as one could alternatively assign the outer oxygen atoms to the linkers.

Especially at high frequencies, the phonon DOSs of all systems are dominated by rather sharp peaks. This suggests mostly weakly dispersing bands. For IRMOF-1(Mg) and IRMOF-14(Mg), the highest-frequency modes occur around 90 THz. These vibrations correspond to C-H stretching motions, as can be concluded from analyzing their eigenvectors. Thus, they do not have equivalents in IRMOF-130, where the linkers do not contain any H atoms. In all systems, the region between 40 and 50 THz is dominated mostly by stretching vibrations of the carbon and oxygen atoms with contributions from hydrogen bending vibrations in IRMOF-1(Mg) and IRMOF-14(Mg). The non-vanishing participation of metal atoms in the higher-frequency vibrations in IRMOF-130(Mg/Ca) is attributed to the very short lateral extent of the linkers in those systems.

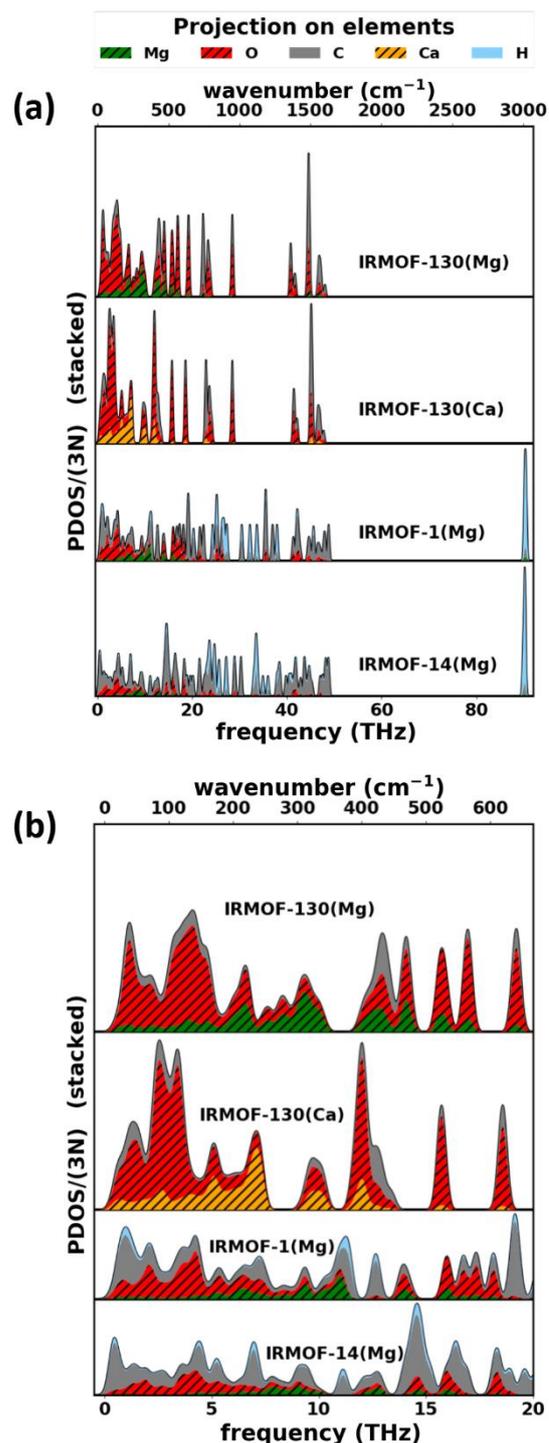

FIG. 3. Projected density of states (PDOS) normalized to the number of bands as a function of frequency for IRMOF-130(Mg), IRMOF-130(Ca), IRMOF-1(Mg), and IRMOF-14Mg). The PDOS contributions per species are plotted in a stacked way so that the outermost line represents the total DOS. Panel (a) shows the entire spectral range, while panel (b) displays a zoom into the low-frequency region up to 20 THz. The hatched contributions stem from elements making up the nodes.

In the region between 29 and 40 THz a wide phonon band gap opens in IRMOF-130(Mg) and IRMOF-130(Ca). Conversely, for IRMOF-1(Mg) and IRMOF-14(Mg) this spectral region is dominated by in-plane bending vibrations of the aromatic linker moieties. Especially in IRMOF-14, the high number of atoms in the linker results in many optical phonon branches almost closing the entire band gap observed in IRMOF-130.

C-H out-of-plane bending modes dominate the next series of bands between 23 and 27 THz in IRMOF-1(Mg) and IRMOF-14(Mg). Some modes in that spectral range also correspond to in-plane deformations of the entire linkers. These are the only modes showing up in that frequency range in IRMOF-130(Mg/Ca).

These observations show that the spectral region above ~20 THz essentially represents vibrations localized in the linkers and the connecting oxygens with only small contributions from the metal atoms.

Below 20 THz, oscillations of the metal atoms increasingly contribute to the phonon density of states and coupled oscillations of linkers and nodes dominate. At very low frequencies (below ~3 THz) the contributions of the metal atoms in IRMOF-1 and IRMOF-14 decrease. This indicates that vibrations in that region are again largely localized in the linkers and oxygens. They, for example, comprise torsional motions of the aromatic systems, as will be discussed in more detail below, when analyzing selected optical phonon bands. The low-frequency modes are particularly relevant, as they are occupied already at low temperatures (at 300 K, $k_BT$ corresponds to ~6 THz).

Interestingly, below a certain frequency [~10.5 THz for IRMOF-130(Mg) and IRMOF-14(Mg); ~7.8 THz for IRMOF-130(Ca); ~12.0 For IRMOF-1(Mg)], there are enough sufficiently dispersing phonon modes to form a continuous DOS without band gaps. This has, for instance, practical implications for phonon transport: a continuous DOS with a multitude of weakly dispersing bands in the same spectral region makes interphonon scattering processes more likely, as it facilitates the simultaneous conservation of energy and momentum in scattering events [53].

The choice of the metal atom contained in the nodes has a particularly large effect in the spectral region up to 20 THz [i.e., the region shown in FIG. 3(b)], as can be inferred from comparing IRMOF-130(Mg) and IRMOF-130(Ca). For the nearly twice as heavy Ca atom in the linkers, the various PDOS features (including the phonon band gaps) shift to distinctly lower frequencies. Such shifts are expected considering the dependence of resonance frequencies on the mass of the oscillating objects. The observation that the shift mostly affects the modes at lower frequencies (up to ~20 THz) is consistent with the above-discussed finding that motions of the metal atoms primarily contribute to vibrations in that spectral region.

Further insight regarding the spectral region most strongly affected by changing the metal atom can be obtained by comparing the (normalized) cumulative DOS (i.e., the integrals over the DOS up to a certain frequency). Fig. 4(a) shows a comparison for IRMOF-130(Mg) and IRMOF-130(Ca). Below ~20 THz the cumulative DOS of IRMOF-130(Ca) is larger than (or equal to) that of its Mg-based analog. This difference essentially disappears for higher frequencies.

The normalized cumulative DOS also provides insight into how the complexity of the linker affects the spectral distribution of phonon modes. Comparing the situations for IRMOF-130 and IRMOF-1(Mg) in Fig. 4(a), one sees that in the latter case the normalized cumulative DOS rises more gradually. This implies that the spectral distribution in IRMOF-1 is shifted to higher frequencies. This observation can be rationalized by the more extended linkers, which give rise to an increasing number of intra-linker modes, which, as discussed above, are typically found at higher frequencies. This trend is further amplified for IRMOF-14.

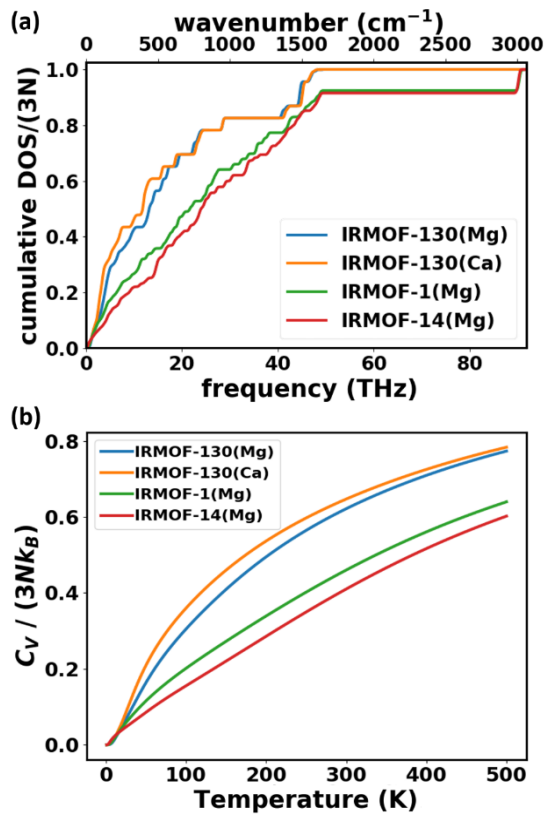

FIG. 4. (a) Normalized cumulative DOS, i.e., DOS integrated up to a given frequency, as a function of that frequency divided by the number of bands. (b) Temperature-dependent phonon heat capacity per unit cell divided by the number of bands.

The main aspects discussed for the cumulative DOS are directly transferred to the (normalized) temperature-dependent heat capacities [Fig. 4(b)] due to the similar nature of the two quantities (for mathematical details see the Supplemental Material [29]). From the heat capacities, one can get an idea of typical temperatures at which the systems behave classically (following the equipartition theorem). Notably, in the displayed temperature range, the normalized heat capacities are still far from unity, i.e., the classical (Dulong-Petit) limit, suggesting that quantum effects in the phonon occupation are still relevant. This trend is much more pronounced for IRMOF-1(Mg) and IRMOF-14 (Mg), consistent with the larger number of high-frequency modes.

## C. Discussion of the (low frequency) band structures

While the DOSs discussed so far are useful for obtaining a general overview of the spectral regions in which certain vibrations occur, further insight can be gained by analyzing the associated phonon band structures. Owing to the huge number of optical phonon modes in all considered MOFs, we will restrict the following discussion to a frequency range of up to 3 THz (i.e., 100 cm$^{-1}$). Even in that region the number of phonon bands is huge, as is shown in FIG. 5. This is a consequence of the significant number of atoms in the primitive unit cells (46 for IRMOF-130, 106 for IRMOF-1, and 190 atoms for IRMOF-14). For IRMOF-1(Mg), we find good qualitative agreement with the band structure of IRMOF-1(Zn) reported in [14].

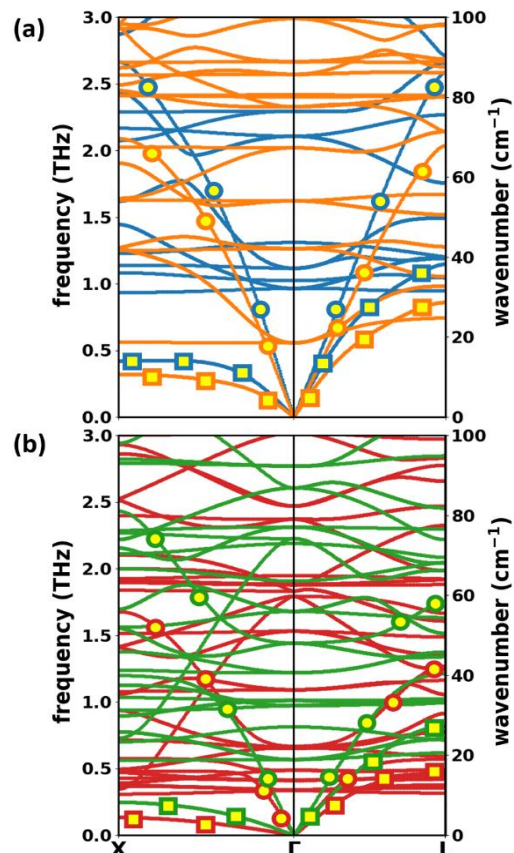

FIG. 5. Phonon band structure of (a) IRMOF-130(Mg) (blue) and IRMOF-130(Ca) (orange) and (b) IRMOF-1(Mg) (green) and IRMOF-14(Mg) (red). Γ-X: Direction corresponding to the linker axis in real space. Γ-L: Direction corresponding to the space diagonals in real space. The round (square) markers show the longitudinal (transverse) acoustic bands.

In the following discussion, we first focus on the acoustic bands. Along the linker direction ($\Gamma$-X), the twofold-degenerate transverse acoustic (TA) bands (the bands lowest in energy marked by squares in FIG. 5) display a significantly smaller dispersion than the longitudinal acoustic (LA) band (marked by circles in FIG. 5). The difference in dispersion can be explained considering the porous structure of the MOFs: For longitudinal acoustic waves all atoms are displaced in a direction of high atomic densities. That is, LA vibrations in the $\Gamma$-X direction correspond to compressions and expansions of the "chain" formed by linkers and nodes. Conversely, TA phonons propagating in $\Gamma$-X cause a displacement of the linkers toward the open pores. This is associated with a smaller restoring force and, consequently, with a reduced band dispersion. This reasoning is also consistent with the observation that the difference between longitudinal and transverse modes decreases along the space diagonal of the pore ($\Gamma$-L). An interesting implication of this finding is that the frequency of the TA modes along $\Gamma$-X should increase upon filling the pores of the MOFs by guest or solvent molecules.

For both IRMOF-130 based systems it is possible to follow the acoustic bands from the $\Gamma$ point to the boundary of the 1BZ [see FIG. 5(a)], which allows us to estimate the corresponding band width. In the $\Gamma$-X direction, for the Mg-based system the width amounts to approximately 2.66 THz for the LA mode and to 0.42 THz for the TA modes. Consistently with the discussion in the previous section, these band widths decrease when replacing Mg with Ca (to 2.08 THz and 0.32 THz, respectively). Along the $\Gamma$-L direction, the band widths decrease slightly for the longitudinal mode [2.58 THz and 2.03 THz for IRMOF-130(Mg) and IRMOF-130(Ca), respectively]. Conversely, for the transverse modes the band widths increase by more than a factor of two compared to $\Gamma$-X (to 1.14 THz and 0.86 THz). This is in line with the above discussion of restoring forces. For IRMOF-1(Mg) and IRMOF-14(Mg) the situation is more involved, as there one encounters avoided band crossings. Analyzing the symmetry of the displacements associated with the phonons at different wavevectors, we are able to identify the parts of the bands with the same character. In Fig. 5(b) the sections corresponding to vibrations with the same character as the longitudinal acoustic bands close to the $\Gamma$-point are highlighted by circles.

Nevertheless, the avoided crossings render a sensible determination of the widths of the LA band in IRMOF-1(Mg) impossible. For the TA bands in the $\Gamma$-X direction avoided crossings cannot occur as this band is the lowest in energy along the entire **q** path. Thus, in this case a band width can be determined (0.25 THz), which is smaller than for both IRMOF-130(Mg) and IRMOF-130(Ca). Not unexpectedly, in IRMOF-14(Mg), with a further increased number of optical bands (already below 0.5 THz), the situation becomes even more complex. Nevertheless, also here a band width for the TA bands along $\Gamma$-X can be determined. With 0.13 THz it is by far the smallest of all considered systems. We attribute this decrease in band width of the TA modes with linker complexity to a decreased resistance of the longer and more flexible linkers to transverse distortions.

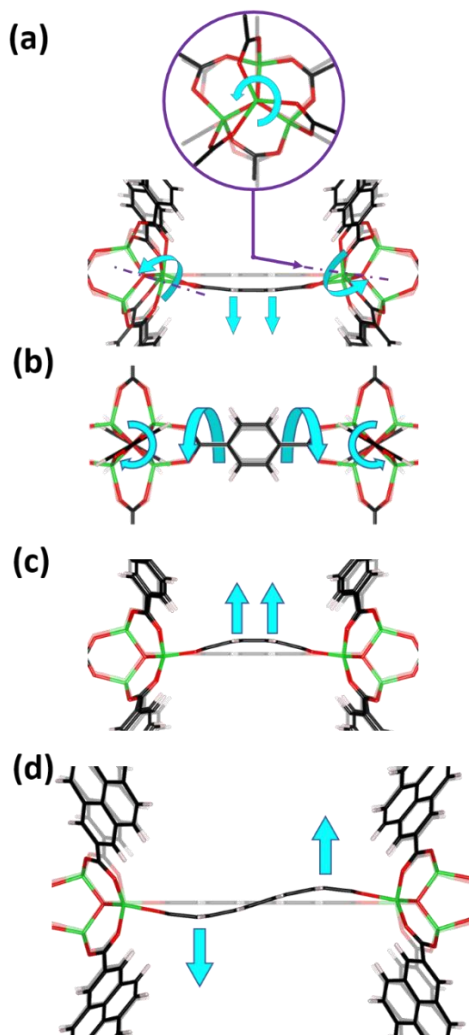

FIG. 6. (a-c) Displacement patterns of low-frequency optical modes in IRMOF-1(Mg). (a) Twisting mode of MgO4 tetrahedra around an axis spanned by the central oxygen atom and the associated magnesium atom at 0.57 THz (19 cm-1) inducing a bending also in the linkers ([110] view; circle: [111] view). (b) Linker torsion around molecular axis at 0.81 THz (27 cm-1) ([100] view), and (c) pure linker bending vibration at 1.44 THz (48 cm-1) ([110] view). (d) Displacement pattern of a second-order bending mode in IRMOF-14(Mg) at 1.79 THz (60 cm-1) ([110] view). The undisplaced geometry is drawn with decreased intensity in the back. Coloring scheme: C, black; Mg, green; O, red; H white. The amplitudes were exaggerated for reasons of clarity. Frequencies are given for the Brillouin zone center.

As far as the optical modes are concerned, their sheer number is too large to discuss all of them in full detail. Therefore, in the following we will describe a few representative examples. In all studied systems, the optical phonon mode with the lowest frequency corresponds to a bending motion of the linkers [see FIG. 6(a)]. In that mode, the metal atoms are essentially frozen while the linker molecules together with the oxygens move. The animation of the corresponding vibration reveals that the $XO_4$ (X = Mg, Ca) tetrahedra twist around an axis through the central O atom of the node and the associated metal atom. To follow that motion of the nodes, the displacements of the atoms in the linkers become particularly large. This explains the predominant linker character of that band in FIG. 7. The low frequency of the mode implies that the restoring force triggered by the vibration is rather weak.

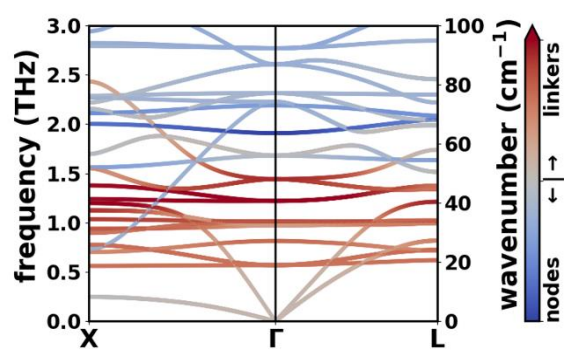

FIG. 7: Phonon band structure of IRMOF-1(Mg) colored according to whether the associated displacements primarily occur in the region of the nodes (metal and oxygen atoms) or linkers (carbons and hydrogens). The mathematical definition of the degree of linker/node localization as well as corresponding plots for all studied MOFs over a wider frequency range can be found in the Supplemental Material [29].

The next mode in IRMOF-1(Mg) occurs at approximately 0.81 THz (27 cm-1; note that here and in the following we refer to frequencies at the Γ-point). The 0.81 THz mode corresponds to a torsional mode, in which the entire aromatic ring undergoes rigid twisting around the long axis of the linker with only little involvement of the oxygens [see FIG. 6(b)]. Correspondingly, as shown in FIG. 7, this mode is again primarily a linker mode. The equivalent mode in IRMOF-14(Mg) occurs already at 0.34 THz (11 cm-1). The significant shift of this mode in the latter system can be explained by the much higher torsional moment of inertia of the more extended linker. This rationale works even quantitatively: the torsional moment of inertia of the twisting moiety (C and H atoms of the linkers) in IRMOF-14 amounts to 509 uÅ$^2$ (with u being the atomic mass unit), while it is only 89.9 uÅ$^2$ for IRMOF-1. As the resonance frequency of a torsional oscillator scales with one over the square root of the moment of inertia, the frequency of the mode should decrease by a factor of 2.38 for IRMOF-14 compared to IRMOF-1. This is exactly the ratio of the torsional frequencies in the full calculations on the two MOFs.

The next higher modes are related to those shown in FIG. 6 (a) and (b), albeit with varying phase shifts between linkers or with displacements affecting only individual linkers. A more detailed discussion of these modes is not instructive, as considering their near degeneracy, it cannot be excluded that different solutions within the resulting "sub-space" of displacements occur upon changing the numerical details of the diagonalization of the dynamical matrix.

The next fundamentally different mode is found at 1.44 THz (48 cm$^{-1}$) in IRMOF-1, and at 0.67 THz (22 cm$^{-1}$) in IRMOF-14. It is characterized by a bending vibration of the linker, as shown in Figure 6(c). As in this mode the oxygens and metal atoms essentially do not move, it displays a nearly exclusive linker-character (see FIG. 7). For the longer linker in IRMOF-14, also a second-order bending vibration is present in the low-frequency region at 1.79 THz (60 cm$^{-1}$). The corresponding displacement pattern is shown in FIG. 6(d). In passing we note that localized vibrations such as those discussed so far are often referred to as rigid unit modes [54], in which entire building blocks of a MOF move rigidly. Rigid unit modes of that kind have been used to describe the elastic behavior [18] or the negative thermal expansion coefficient in IRMOFs [10]–[13],[15] (in addition to mechanism related to acoustic phonon softening [14]).

Another type of mode that should be mentioned here (as will become relevant for the later discussion) is characterized by a torsional motion of the carboxylic oxygens around the linker axis (FIG. 8). In IRMOF-1 and IRMOF-14 such modes occur at essentially the same frequency at Γ (1.91 THz = 64 cm$^{-1}$ in IRMOF-1, 1.95 THz = 65 cm$^{-1}$ in IRMOF-14). In line with the displacement pattern in FIG. 8, these modes display a dominant node character (see FIG. 7). In IRMOF-130(Mg) and IRMOF-130(Ca) the corresponding modes are found at 1.31 THz and 2.41 THz, respectively. A variant of these modes in which also the metal atoms bonded to the carboxylic oxygens participate in the torsional motion is found at 2.30 THz in IRMOF-130(Mg) and at 2.02 THz in IRMOF-130(Ca). Interestingly, the frequencies of the former modes are higher in the Ca-based than in the Mg-based MOF, although typically the larger mass of the metal atom results in a shift of the peaks in the opposite direction. We explain this behavior for the torsional motion of the oxygens by the steeper torsional potential we calculate for the Ca-based system (see Supplemental Material [29]); i.e., this effect is not related to the mass of the metal atoms but rather to their bonding properties.

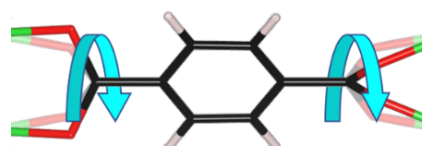

**FIG. 8.** Displacement pattern of a typical optical mode in which the carboxylic oxygens oscillate around the linker axis, while the rest of the system does not move. Such modes occur in all systems [e.g., at 1.91 THz (64 cm-1) in IRMOF-1(Mg)] at Γ ([100] view). The undisplaced geometry is drawn with decreased intensity in the back. Coloring scheme: C, black; Mg, green; O, red; H white. The amplitudes were exaggerated for reasons of clarity.

### D. Phonon group velocities

An important phonon property that can be directly extracted from the gradient of the band structures is the group velocity. It is of distinct relevance for many transport processes, as it quantifies the speed at which thermal energy is transported by a given phonon. Coloring the phonon bands of the investigated systems according to their group velocities (FIG. 9), it becomes apparent that longitudinal acoustic modes have by far the highest group velocities in the low-frequency region. The observation that the group velocities for transverse acoustic modes are significantly smaller is consistent with the notion that in the studied MOFs resistance to transverse deformation is much weaker than resistance to longitudinal compression. As indicated already in Sec. IV C, this behavior is particularly pronounced for modes propagating in the Γ-X direction. This indicates that shear waves propagating along the linker direction experience unusually weak restoring forces due to the porous nature of MOFs.

To assess the role played by the specific structure of the MOFs, as a first step, we analyze the group velocities of the acoustic modes in the long-wavelength limit (i.e., for $\mathbf{q}\to 0$). They correspond to the respective speeds of sound, c. Interestingly, as shown in FIG. 10(a), we find that for the majority of the propagation and displacement directions the values of c only weakly depend on the linker [see, e.g., slopes of the fitted straight lines listed left of the data points in FIG.10(a)]. Note that the speed of sound in FIG. 10 is plotted over the reciprocal lattice constant. The reason for this will become clear in the following discussion.

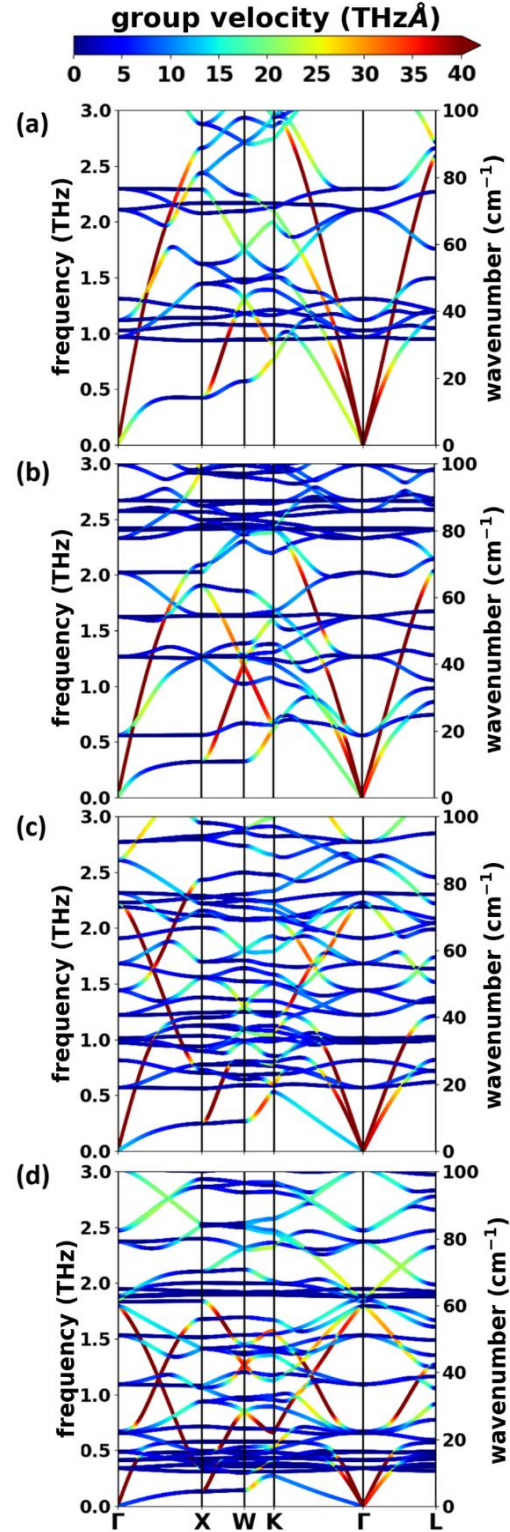

**FIG. 9:** Phonon band structures of (a) IRMOF-130(Mg), (b) IRMOF-130(Ca), (c) IRMOF-1(Mg), and (d) IRMOF-14(Mg) zoomed in the very low frequency region, colored according to the norm of the group velocities. The group velocity scale was truncated at 40 THzÅ for reasons of visibility. Γ-X: Direction corresponding to the linker axis in real space. Γ-K: Direction corresponding to the face diagonals in real space. Γ-L: Direction corresponding to the space diagonals in real space.

Heavier metals in the node (Ca versus Mg) somewhat reduce $c$, but the effect is not particularly strong. The only two types of modes for which $c$ significantly depends on the structure of the MOF are the (degenerate) transverse acoustic modes in the $\Gamma$-X directions and one of the transverse acoustic modes in the $\Gamma$-K direction (the one characterized by a displacement perpendicular to the face of the cube within which the wave propagates).

To rationalize the only very weak dependence of the speed of sound on the MOF structure, it is useful to consider the example of a 1D chain of masses, $m$, separated by a distance, $a$, and elastically coupled by a force constant, $\gamma$. In this case, one obtains the following expression for $c$:

$$c = a\sqrt{\frac{\gamma}{m}} \qquad (3)$$

A similar relation holds for cubic crystals along high-symmetry directions [55]. Multiplying Eq. (3) by the mass density, $\rho = m/a^3$, one obtains:

$$c\rho = \sqrt{\gamma m}\,\frac{1}{a^2} \qquad (4)$$

Interestingly, when plotting $c\rho$ as a function of $a^{-2}$, or the respective square roots [see FIG. 10(b)], one obtains an essentially linear relationship. In the spirit of Eq. (4), this suggests that the product of stiffness times inertia, $(\gamma m)^{1/2}$, is constant throughout all studied systems. In other words, increasing the size of the linker changes the mass of the unit cell, but simultaneously decreases the effective stiffness of the system. Consequently, in the argument of the square root of Eq. (3), the numerator decreases while the denominator increases with the size of the linker. Longer linkers, however, also result in an increase of the unit-cell length, $a$, which cancels the decrease in $(\gamma/m)^{1/2}$. The compensation works best for modes with a speed of sound around 6000 ms$^{-1}$ [as exemplified by the slopes k of the linear fits in FIG. 10(a)]. For modes with a smaller value of $c$, the speed of sound decreases with linker length, where the magnitude of the effect is amplified for larger deviations from $c \approx 6000$ ms$^{-1}$.

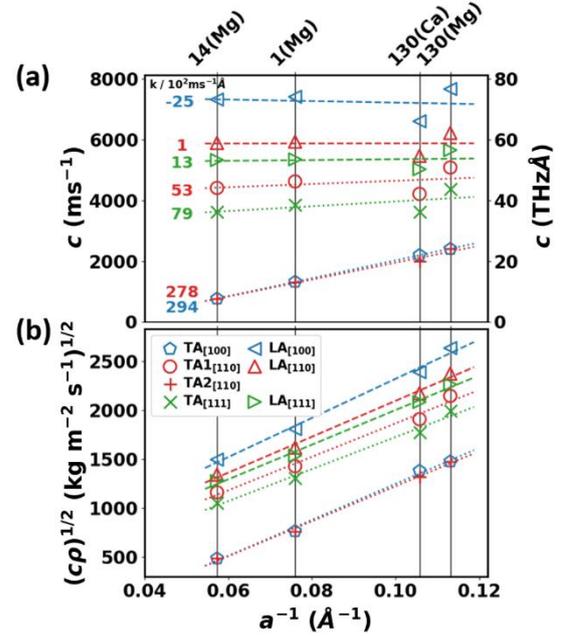

FIG. 10. (a) Dependence of the speed of sound for different directions and polarizations on the MOF structure. For consistency with panel (b) the structure of the MOF is characterized by the inverse unit cell length. The straight lines are linear fits through the respective data points. The slopes, k, of those fits are given left of the lines in units of 10$^2$ms$^{-1}$Å$^{-1}$. An equivalent plot (and equivalent fits) for a linear length scale can be found in the Supplemental Material [29]. Notably, whether c is plotted as a function of a or of a$^{-1}$ does not qualitatively change the situation. (b) Square roots of group velocities times mass densities as a function of a$^{-1}$ for the same modes as in panel (a). The dashed and dotted straight lines are again linear fits. The pentagon, circle, and two crosses denote the transverse acoustic modes, where the two modes are degenerate for the $\Gamma$-X and $\Gamma$-L directions. The different triangles denote longitudinal acoustic modes.

Regarding the group velocities of the optical modes, the situation is again rather complex due to their high number. Thus, to analyze the overall situation, we resort to densities of states in analogy to Sec. IV B. Correspondingly, FIG. 11 shows such densities of states, which now are resolved not only with respect to $\omega$ but also with respect to the norm of the group velocity vector (see Sec. III C; the scatter plots of $\|\mathbf{v_g}\|$ as a function of the frequency, from which these densities can be derived, are shown in the Supplemental Material [29]).

Several conclusions can be drawn from these densities:

(i) The modes with the highest group velocities for all MOFs are found below 3 THz. The associated band diagrams (see FIG. 9) reveal that they can be associated with the longitudinal acoustic phonon branches. In that region, group velocities beyond 60 THzÅ (i.e. 6000 ms$^{-1}$) are reached. These values are comparable to those observed in many nonporous materials, which is consistent with the finding that the porosity of the materials primarily impacts the transverse modes.

(ii) Interestingly, in all systems also in the region of the optical phonons, one observes group velocities amounting to several tens of THzÅ. In fact, for IRMOF-1 and IRMOF-14 there is a pronounced peak around 10 THz. Such "fast" optical phonons will be thermally occupied at room temperature (at 300K 10 THz corresponds to ~1.6 $k_B T$). This suggests that their contribution to processes such as thermal transport can be significant. Notably, in all systems there are also significantly dispersing bands between 40 THz and 50 THz, where we find the backbone stretching vibrations of the linkers.

(iii) In spite of the very high group velocities observed in FIG. 11, the densities of the associated phonons are several orders of magnitude smaller than those of phonons with group velocities of only a few THzÅ. This is not well resolved for the logarithmic density scales chosen for all panels in FIG. 11. Therefore, in FIG. 12, we show the normalized density of group velocities in a linear plot. That quantity is obtained by integrating the data from FIG. 11 over all frequencies (see Supplemental Material [29]).

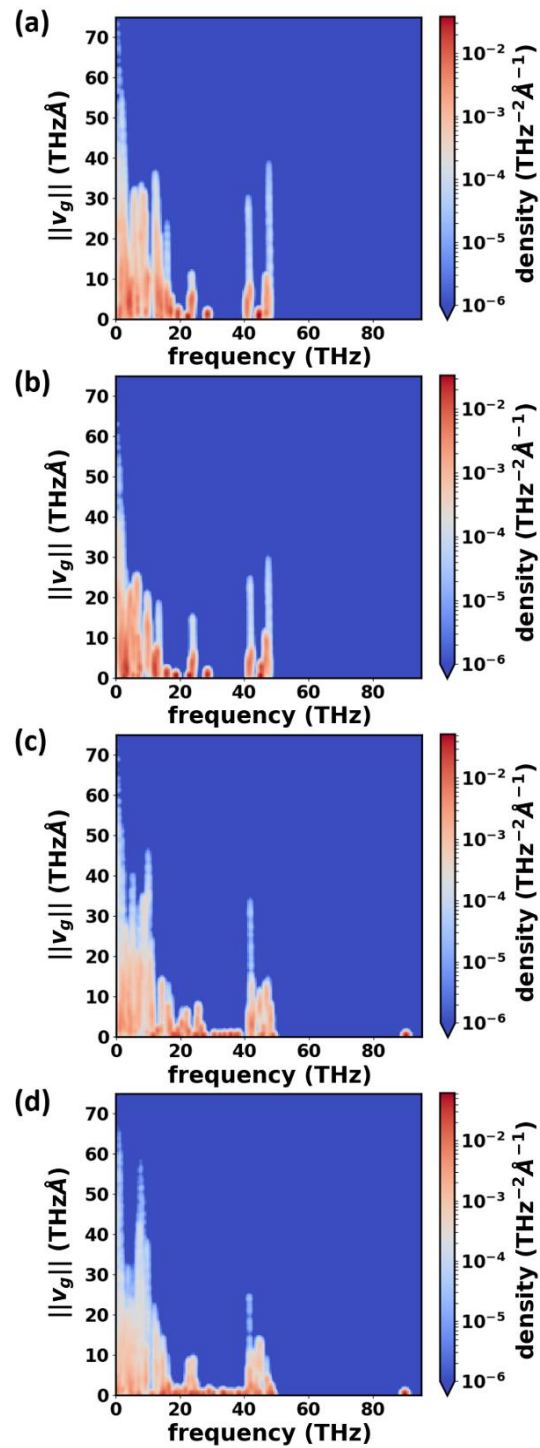

FIG. 11. Normalized frequency and group velocity resolved DOS of (a) IRMOF-130(Mg), (b) IRMOF-130(Ca), (c) IRMOF-1(Mg), and (d) IRMOF-14(Mg). A high density indicates that many modes with that frequency and group velocity can be found in the evenly sampled 1BZ.

(iv) The data in FIG. 12 support the qualitative statements from (i)-(iii). They additionally show that upon increasing the length of the linkers, modes with particularly small group velocities increasingly dominate. Moreover, a comparison between IRMOF-130 (Mg) and IRMOF-130(Ca) shows that also increasing the mass of the metal atoms reduces the most frequent group velocity.

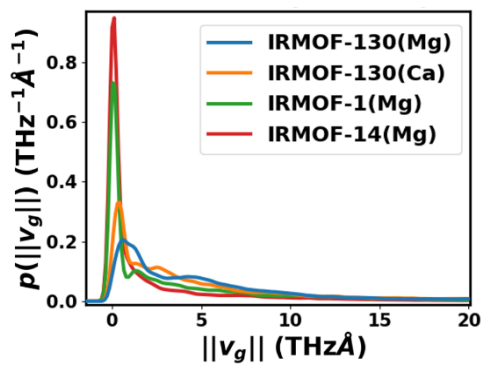

**FIG. 12.** Normalized density of group velocities for the studied systems.

### E. Correlation between mode localization and group velocities

Considering the heterogeneous structure of MOFs, different optical modes are often localized largely on the nodes or largely on the linkers (see discussion in section IV.C). This raises the question of whether there is a direct correlation between the degree of localization of a mode and its group velocity. To analyze that, it is necessary to first quantify the degree of localization. This can be done via the mode-participation ratios, PRs, defined in Sec. III C, where a participation ratio of 1 denotes a fully delocalized mode, while small participation ratios denote strong localization. Phonon band structures colored according to the mode participation ratios are shown in FIG. 13 for all investigated systems along a continuous path of high-symmetry directions.

A general observation is that the strongly dispersing bands with large group velocities (cf. FIG. 8) are characterized by high PRs with essentially all atoms participating in the respective oscillations. The only rather strongly dispersing band with a comparably small participation ratio is the LA band in the $\Gamma$-K direction in IRMOF-130 (Mg/Ca) in the vicinity of the avoided crossing between 1.0 THz and 1.5 THz.

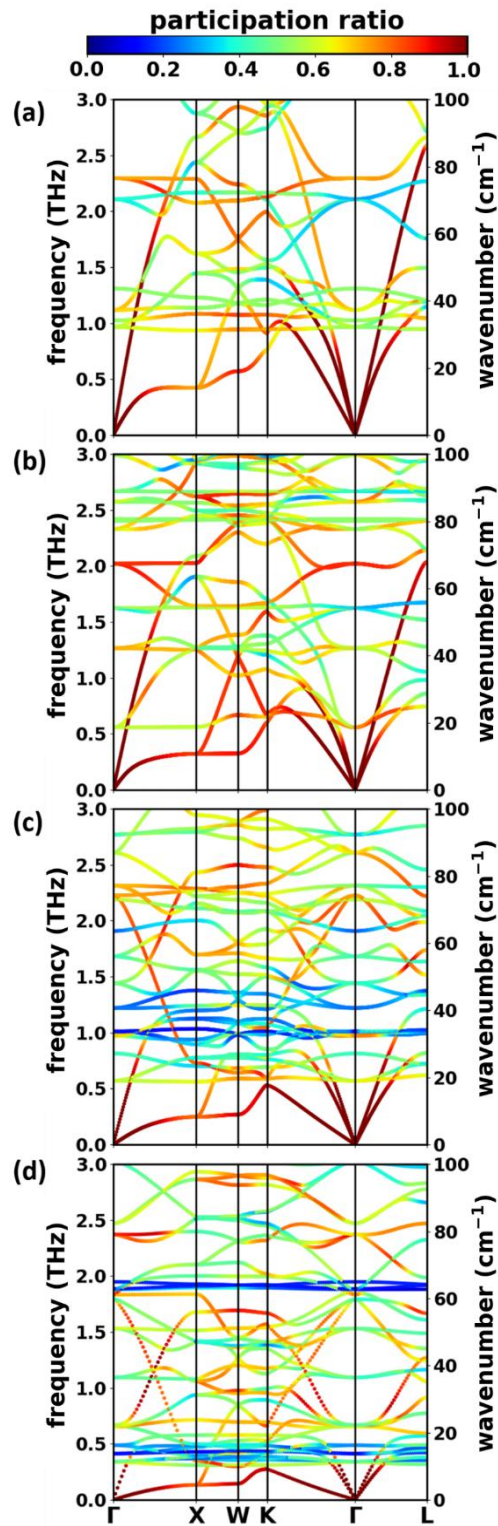

**FIG. 13.** Phonon band structures of (a) IRMOF-130(Mg) (b) IRMOF-130(Ca), (c) IRMOF-1(Mg), and (d) IRMOF-14(Mg) zoomed in the very low frequency region, colored according to the mode participation ratios. $\Gamma$-X: Direction corresponding to the linker axis in real space. $\Gamma$-K: Direction corresponding to the face diagonals in real space. $\Gamma$-L: Direction corresponding to the space diagonals in real space.

16 / 23

Notably, the opposite correlation does not hold; i.e., there are multiple bands, which are essentially flat (at least for a certain range of wave vectors), but for which the PRs are large. Close to the Brillouin-zone boundaries, this can be understood from the "standing-wave" nature of the corresponding oscillations. This explanation, however, does not apply for instance, to the flat band at 2.3 THz (2.0 THz) in IRMOF-130(Mg) [IRMOF-130(Ca)], which has a PR of 0.72 (0.86). The reason for that lies in the nature of the atomic displacements associated with these modes. As mentioned in Sec. IV C, they correspond to torsional vibrations of the carboxylic oxygens together with the atoms in the metal linkers. Due to the short linkers, this involves nearly all atoms of the MOF, but in the absence of hydrogen atoms in the linker, the torque generated by the torsion is not transmitted through the carbon single bond that connects two adjacent nodes. Thus, a phase shift between torsional motions of carboxylic groups in neighboring nodes hardly affects the phonon energy resulting in flat bands.

As far as the other phonon modes discussed in Sec. IV C are concerned, the values of their participation ratios can be well understood from the corresponding displacement patterns. To more quantitatively correlate the mode participation ratio and the group velocity, we plot the data for all phonons coinciding with the **q**-grid specified in the figure caption for the most "extreme" cases [IRMOF-130(Mg) and IRMOF-14(Mg)]. The data for the other two systems are contained in the Supplemental Material [29]. The frequency of each mode is included via a color code.

FIG. 14 further supports the notion that there is no direct one-to-one correspondence between the localization of a vibration and the group velocity of the associated phonon (as for such a direct correlation all points would be grouped around a straight line through the origin). Rather the vast majority of the phonons have a group velocity below 20 THzÅ (consistently with FIG. 12) and participation ratios between 0.1 and 0.8.

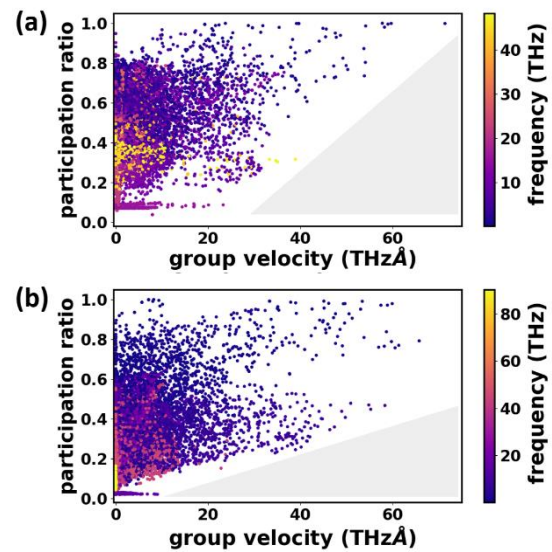

FIG. 14. Correlation between mode participation ratio and norm of the group velocity for all phonons considered in the determination of the densities of states. The frequency of each phonon is included via a color code. (a) IRMOF-130(Mg) (calculated on a 12×12×12 *q* mesh), and (b) IRMOF-14(Mg) (calculated on a 10×10×10 *q* mesh). The gray triangles are a guide to the eye denoting a region with no modes.

FIG. 14, however, also shows that if the participation ratio is small, this limits the achievable group velocity (see gray-shaded triangles in FIG. 14), which is again consistent with the conclusions drawn above from the band structures.

In passing we note that one could also wonder whether there was a correlation between the group velocities and the degree to which atoms in the unit cell move in the same direction. Therefore, as discussed in the Supplemental Material [29], we defined a corresponding parameter in analogy to the mode participation ratio. Large values of that "acoustic participation ratio" were, however, obtained only for some of the acoustic modes and there was no correlation between that parameter and high group velocities in optical bands.

## F. Anharmonic properties

As a last aspect we briefly address anharmonicities in the studied systems, where an in-depth investigation would lie beyond the scope of this article. Anharmonic properties are relevant insofar as they determine processes such as thermal expansion and heat transport (the latter via phonon lifetimes). A consequence of anharmonicities in the potential-energy surface is also that (within the quasi-harmonic approximation) phonon frequencies change with the volume of the unit cell. Knowing the magnitude of that effect allows us to study thermodynamic properties as a function of pressure.

The relevant quantities in this context are the mode Grüneisen parameters $\gamma_{q,n}$, which relate the observed relative frequency change, $d\omega_{q,n}/\omega_{q,n}$, to the relative change in unit cell volume, $dV/V$ ($\gamma_{q,n}=-V/\omega_{q,n} \times d\omega_{q,n}/dV$). Scatter plots of the mode Grüneisen parameters as a function of the associated frequencies are shown in FIG. 15(a) (note the logarithmic frequency scale emphasizing the low-frequency situation).

Notably, independently of the considered MOF, most of the mode Grüneisen parameters are close to zero for frequencies above ~3 THz. Unlike in most materials, the majority of the mode Grüneisen parameters of the systems studied here have a negative sign. This means that phonon frequencies decrease with decreasing volume (phonon softening), which leads to negative thermal expansion coefficients. This property is, indeed, frequently observed in certain MOFs [10]–[15],[21].

The magnitude of the mode Grüneisen parameters is smallest in IRMOF-1(Mg). Changing the linkers to either IRMOF-130(Mg) or IRMOF-14(Mg) increases the maximum values of $\gamma_{q,n}$. The maximum mode Grüneisen parameters also significantly increase upon replacing Mg with Ca.

Plotting the band structures colored according to the Grüneisen parameters (see Supplemental Material [29]) reveals that by far most negative values of $\gamma_{q,n}$ are obtained for the TA bands along $\Gamma$-X-W and for the lowest-lying optical mode [characterized by oxygen atoms rotating around the metal inducing a pronounced linker bending; see FIG. 6(a)]. Indeed, these modes have already been identified as being a possible reason for the negative thermal expansion of IRMOF-1 in the past [11]–[15].

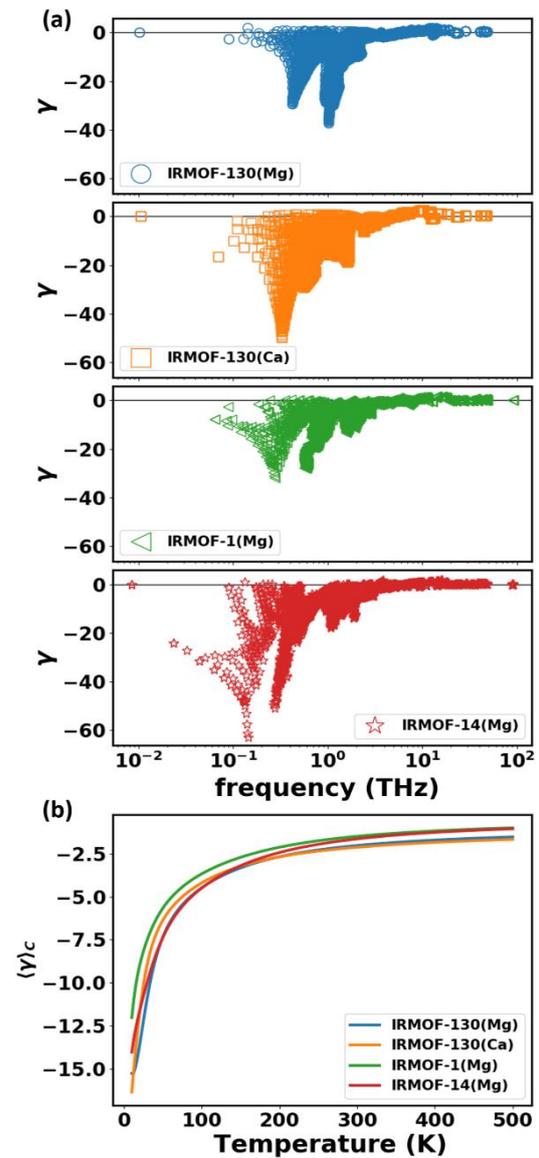

**FIG. 15:** (a) Scatter plot of mode Grüneisen parameters, $\gamma$, as a function of frequency (logarithmic scale) calculated on a 20x20x20 $q$ mesh. (b) Temperature-dependence of mean Grüneisen parameter $\langle\gamma\rangle_C$ averaged with the heat capacity.

As far as possible correlations between the Grüneisen constants and other vibrational properties are concerned, we observe the following: (i) There is no direct correlation between $\gamma_{q,n}$ and the participation ratios, but the highest values for $\gamma_{q,n}$ are usually found for modes with high PRs. (ii) Large negative Grüneisen parameters are obtained mostly for modes with small group velocities (i.e., flat bands). The corresponding data are shown in the Supplemental Material [29].

A relevant derived quantity is the temperature-dependent mean Grüneisen parameter, $\langle\gamma\rangle_C$, which is intimately related to thermal expansion. It can be calculated from the mode Grüneisen parameters weighting them by the temperature-dependent mode contribution to the heat capacities normalized by the total heat capacity (see Supplemental Material [29]). $\langle\gamma\rangle_C$ can be thought of as an average Grüneisen parameter considering thermal occupation and the phonon DOS. FIG. 15(b) shows that $\langle\gamma\rangle_C$ is less system dependent than the mode Grüneisen parameters, which is largely due to the maxima in $\gamma_{q,n}$ appearing at different frequencies and a compensation of the system-dependence of the values of $\gamma_{q,n}$ and the heat capacities.

## V. SUMMARY AND CONCLUSION

We have analyzed the phonon properties of several prototypical isoreticular metal-organic frameworks. They turn out to be highly complex owing to the sheer number of phonon bands (138 bands even in the smallest considered system, IRMOF-130). Nevertheless, several general trends can be identified: the high-frequency region of the density of states is dominated by vibrations largely localized on the linker parts of the MOFs. Only for frequencies below ~20 THz, oscillations of the metal-oxide nodes start contributing significantly. As a consequence, the spectral weight of the density of states shifts to higher frequencies for MOFs with more extended linkers. Notably for the latter systems (i.e., IRMOF-1 and IRMOF-14), the spectral region below ~ 3 THz is again dominated by vibrations involving primarily the linkers.

As far as the acoustic phonon bands are concerned, there are large differences in the band dispersions and group velocities between transverse and longitudinal modes, especially for propagation in the $\Gamma$-X direction (i.e., parallel to the linkers). This can be rationalized by the porous structure of the MOFs, which results in comparably weak restoring forces upon displacements perpendicular to the linker axes. Conversely, restoring forces for longitudinal acoustic displacements are sizable, as can be inferred from the rather large associated group velocities for long wavelengths. These, for the studied systems, range between 5000 ms$^{-1}$ and 7700 ms$^{-1}$, comparable to and even exceeding many densely packed solids. Notably, for the considered systems these group velocities for a given high-symmetry direction are largely independent of the chosen linker.

As far as the optical modes are concerned, even in the low-frequency region (up to ~3 THz) we observe a variety of different vibrations ranging from modes dominated by twisting motions of the nodes and torsional vibrations of the linkers, via linker bending modes of different order, to twisting modes of the carboxylic groups. Here we observe relations known from classical mechanics, like a proportionality of the frequencies of the linker torsional motion to one over the square root of the torsional moment of inertia.

An observation relevant for all processes involving phonon transport in MOFs is that several of the optical phonons have sizable group velocities, although the highest group velocities are still observed for the longitudinal acoustic bands. With increasing linker complexity, one observes that the density of states with large group velocity diminishes, and phonons with small group velocities become more abundant.

An interesting question for heterogeneous systems such as MOFs is whether there is a correlation between the (de)localization of a vibration (expressed via the mode participation ratio) and the group velocity of the associated phonon. Generally, we find that a one-to-one correspondence between these two quantities does not exist. Nevertheless, a sufficiently large mode participation ratio is apparently a prerequisite for a large group velocity. The opposite, namely that group velocities for delocalized vibrations are always large is, however, not observed. The reason for this is that for certain displacement patterns, the couplings between neighboring unit cells remain small, even if virtually all atoms of the MOF participate in the vibration.

Regarding the role of anharmonicities, we observe pronounced phonon softening (i.e., an increase of phonon frequencies with increasing volume, which typically results in negative thermal expansion coefficients). Sizable negative mode Grüneisen parameters are, however, observed only for frequencies below ~3 THz, where the largest values are found for transverse acoustic phonons and the lowest-lying optical modes. Interestingly, anharmonicities are smallest for IRMOF-1(Mg), while changing the linker or increasing the mass of the metal atoms from Mg to Ca increases anharmonic effects.

These results highlight the rather complex correlation between MOF structure and phonon properties. They also show that promising properties, like high group velocities for acoustic as well as optical phonons, are well within reach in spite of the heterogeneous nature of the MOFs so that phonon engineering could soon yield materials with tailored properties for specific applications.

## ACKNOWLEDGMENTS

T.K. and E.Z. acknowledge the Graz University of Technology for financial support through the Lead Project (LP-03) and the use of HPC resources provided by the ZID. The computational results have been in part achieved by using the Vienna Scientific Cluster (VSC3). N.B. gratefully acknowledges financial support under the COMET program within the K2 Center "Integrated Computational Material, Process and Product Engineering (IC-MPPE)" (Project No. 859480). This program is supported by the Austrian Federal Ministries for Transport, Innovation, and Technology (BMVIT) and for Digital and Economic Affairs (BMDW), represented by the Austrian research funding association (FFG), and the federal states of Styria, Upper Austria, and Tyrol. This work has partially been funded by the Austrian Research Promotion Agency through the project ThermOLED (FFG No. 848905).## References

[1] J. Lee, O.K. Farha, J. Roberts, K.A. Scheidt, S.T. Nguyen, and J.T. Hupp, Metal-organic framework materials as catalysts, Chem. Soc. Rev. **38**, 1450 (2009).
[2] G. Ferey, Hybrid porous solids: past, present, future., Chem. Soc. Rev. **37**, 191 (2008).
[3] L.J. Murray, M. Dinca, and J.R. Long, Hydrogen storage in metal-organic frameworks., Chem. Soc. Rev. **38**, 1294 (2009).
[4] S. Keskin, T.M. van Heest, and D.S. Sholl, Can metal-organic framework materials play a useful role in large-scale carbon dioxide separations?, ChemSusChem **3**, 879 (2010).
[5] T.A. Makal, J.R. Li, W. Lu, and H.C. Zhou, Methane storage in advanced porous materials, Chem. Soc. Rev. **41**, 7761 (2012).
[6] S. Kitagawa, R. Kitaura, and S. Noro, Functional porous coordination polymers., Angew. Chem. Int. Ed. Engl. **43**, 2334 (2004).
[7] A.M. Walker, B. Civalleri, B. Slater, C. Mellot-Draznieks, F. Corà, C.M. Zicovich-Wilson, G. Román-Pérez, J.M. Soler, and J.D. Gale, Flexibility in a metal-organic framework material controlled by weak dispersion forces: The bistability of MIL-53(Al), Angew. Chemie - Int. Ed. **49**, 7501 (2010).
[8] G. Kieslich, S. Kumagai, K.T. Butler, T. Okamura, C.H. Hendon, S. Sun, M. Yamashita, A. Walsh, and A.K. Cheetham, Role of entropic effects in controlling the polymorphism in formate ABX3 metal-organic frameworks, Chem. Commun. **51**,